\documentstyle[12pt]{article}

\begin{document}
\def\ra{\rightarrow}
\def\ie{\hbox{\it i.e.}}        \def\etc{\hbox{\it etc.}}
\def\eg{\hbox{\it e.g.}}        \def\cf{\hbox{\it cf.}}
\def\VEV#1{\left\langle #1\right\rangle}
\let\vev\VEV
\def\half{{1\over 2}}
\def\beq{\begin{equation}}
\def\eeq{\end{equation}}
\def\bea{\begin{eqnarray}}
\def\eea{\end{eqnarray}}
\begin{titlepage}
\baselineskip 0.35cm
\begin{center}
\today
\hfill SNUTP 92-87\\
\null\hfill \\
\vskip 1.5cm
{\large \bf   Constraints from
Nucleosynthesis and SN1987A }
\\
{\large \bf
on  Majoron Emitting  Double Beta Decay}
\vskip 1.7cm
{Sanghyeon Chang}
\vskip .05cm
{ Department of Physics and Center for Theoretical Physics}
\vskip .05cm
{Seoul National University, Seoul, 151-742 Korea}
\vskip 0.5cm
{Kiwoon Choi}
\vskip .05cm
{ Department of Physics, Chonbug National University}
\vskip 0.05cm
{Chonju, 560-756 Korea}

\vskip 1.5cm

{\bf Abstract}
\end{center}
\begin{quotation}
\baselineskip .73cm
We examine whether observable majoron emission in double beta decay
can be compatible with the big-bang  nucleosynthesis (NS)
and the observed neutrino flux from SN1987A.
It is found that
 the NS upper bound on $^4$He abundance implies  that
 the majoron-neutrino Yukawa coupling constant   $g\leq 9\times 10^{-6}$
 and its maximal value is
allowed only when the scalar quartic coupling constant $\lambda$
is extremely small, $\lambda\leq 100g^2$.
 It is also observed that, although quite less restrictive,
SN1987A also provides independent  constraints  on coupling constants.

\end{quotation}
\baselineskip .75cm
\end{titlepage}
\setcounter{page}{2}

Majoron is the massless Goldstone boson
associated with    spontaneous lepton number violation $\cite{majoron}$.
If exists, it would lead to many interesting phenomenological
consequences.
Amongst them, one particularly interesting
phenomenon is  ``majoron emitting neutrinoless double beta decay"
($\beta\beta_J$) $\cite{beta}$.
Recently it was pointed out that the potential anomaly in the
double beta decay spectra of several elements may be explained by
$\beta\beta_J$ $\cite{cline}$. The desired
value of the Yukawa coupling constant
is roughly $g_{ee}\simeq 10^{-4}$.
Regardless of  this observation,
if $g_{ee}$ is not very small, \eg \, not less than $10^{-5}$,
one may be able to observe $\beta\beta_J$ in the near future.
As was shown in ref. $\cite{valle}$,
it is not difficult to construct majoron models  which
provides such a  value of $g_{ee}$  while
satisfying all known experimental constraints.
However in view of the strong cosmological and astrophysical
implications of majoron $\cite{kim}$, it is desirable to examine
whether observable $\beta\beta_J$ can be compatible
with the big-bang cosmology and also with astrophysical observations.
In this paper, we consider possible constraints from
the big-bang nucleosynthesis  and the supernova
 SN1987A on majoron models
in which observable $\beta\beta_J$ is possible.

If the
lepton number symmetry $L$ is spontaneously broken at an energy  scale $v$,
 majoron-neutrino Yukawa coupling
matrix $g_{\alpha\beta}$ ($\alpha$,$\beta$=$e,\mu,\tau$)
 is related to the neutrino mass matrix $m_{\alpha\beta}$
as $g_{\alpha\beta}=m_{\alpha\beta}/v$.
Current data on $\nu$-less $\beta\beta$ decay implies
$m_{ee}\leq 1$ eV and $g_{ee}\leq {\rm few}\times 10^{-4}$
with  uncertainties arising from   nuclear matrix elements
$\cite{beta}$.
For $\beta\beta_J$ to be observable in the near  future,
 we may need $g_{ee}\geq 10^{-5}$.
This implies that $L$-breaking scale is
very  small compared to the Fermi scale,
\beq
v\leq 100 \, \, {\rm keV},
\eeq
leading to  a  fine tuning problem in general $\cite{foot1}$.
Here we will  not concern this theoretical difficulty.
We rather concentrate on models with such a low $L$-breaking scale
 to see  whether observable $\beta\beta_J$ can be compatible with
the standard nucleosynthesis model $\cite{ktbook}$ and the observed neutrino
flux from SN1987A $\cite{1987a}$.

In  majoron models adopting low energy $L$-violation (1) to provide
observable $\beta\beta_J$,
low energy  majoron-neutrino interactions
can be described by  the effective lagrangian:
\bea
{\cal L}_{eff}&=&\bar{\nu}_{\alpha}i\gamma\!\cdot\!\partial\nu_{\alpha}
+|\partial_{\mu}\chi|^2-
\frac{1}{\sqrt{2}}(g_{\alpha\beta}\chi^{\ast}\bar{\nu}^c_{\alpha}\nu_{\beta}
+{\rm h.c.}) \nonumber \\
&&-\lambda (\chi\chi^{\ast}-\frac{v^2}{2})^2+...
\eea
where $\chi$ is a (mostly) gauge singlet Higgs field carrying
the lepton number two,
 and the ellipsis denotes generic nonrenormalizable interactions
suppressed by the powers of  cutoff  scale $\Lambda$ which is
usually taken to be the Fermi scale.
 The majoron field $J$  appears in $\chi$ as
\beq
\chi =\frac{1}{\sqrt{2}}(v+\rho+iJ),
\eeq
and the mass eigenstates neutrino $\nu_i=\sum_{\alpha}
U_{i \alpha}\nu_{\alpha}$
 ($i=1,2,3$) has the Yukawa interaction with $J$ and $\rho$
 whose coupling constant is given by $g_i=|\sum_{\alpha\beta}
 U^{\ast}_{i\alpha}U^{\ast}_
 {i\beta}g_{\alpha\beta}|$ for a unitary mixing matrix $U_{i\alpha}$.

Let us now consider possible constraints on
majoron-neutrino interactions  from the big-bang nucleosynthesis
(NS) $\cite{bertolini}$.   As is well known, in the standard model of NS,
 the energy density of exotic  particles
at the NS epoch ($T_{NS}\simeq 1$ MeV) is severely constrained
$\cite{ns}$.
In the case that $L$ is restored
and also there is no sizable lepton number excess,
  the observed $^4$He abundance implies
\beq
Y(T_{NS})\equiv \rho_{\chi}(T_{NS})/\rho_{\nu}(T_{NS})\leq 0.3,
\eeq
where $\rho_{\chi}$ and $\rho_{\nu}$ denote the energy
density of $\chi$ and a single species of left-handed
neutrino respectively $\cite{enqvist}$.
 Since $v< T_{NS}$, before  the NS
 the ratio of the Hubble expansion rate
$H$ to the $\nu$-$\chi$  interaction rate  behaves as
$\Gamma_{\rm int}/H\sim T^{-1}$.
Then $\chi$'s would be
rare at high temperature, but eventually enter into a
thermal equilibrium with neutrinos at some temperature
$T_{\rm eq}$ at which $\Gamma_{\rm int}\simeq H$.
For the NS constraint (4) to be satisfied,
we need  $T_{\rm eq}< T_{NS}$.  Here we will
directly  evaluate $\rho_{\chi}$  and  apply the constraint (4),
rather than using more naive condition $T_{\rm eq}< T_{NS}$.

To evaluate $\rho_{\chi}$, we first need to know
whether $L$ is restored around the NS epoch.
The effective mass of $\chi$ in the early universe
is  given by $\cite{finitetem}$
 \bea
 m^2_{\rm eff}&=
 &-\lambda v^2+\int \frac{d^3 k}{(2\pi)^3 E}(4\lambda f_{\chi}
 +2\, {\rm tr}(gg^{\dagger}) f_{\nu})   \nonumber \\
 &=&-\lambda v^2+
  4\lambda n_{\chi}\vev{1/E_{\chi}}+ {\rm tr}(gg^{\dagger}) T^2/12,
 \eea
where $f_X$ ($X=\chi,\nu$) denotes the phase space distribution
function of $X$ whose  number density is
defined as $n_X=\int d^3 k \, f_X/(2\pi)^3$, and
 neutrinos are assumed to be in thermal equilibrium.
To proceed, let us set
\beq
n_{\chi}\vev{1/E_{\chi}}=\xi Y^{\omega} n_{\nu}\vev{1/E_{\nu}}=\xi Y^{\omega}
T^2/24,
\eeq
where $Y=\rho_{\chi}/\rho_{\nu}$.
Clearly  the average energy of $\chi$
do {\it not} exceed that of $\nu$ and thus $n_{\chi}/n_{\nu}\geq
Y$,  implying  that $\xi Y^{\omega}\geq Y$.
Then for $v\leq 100$ keV and
 $Y$ saturating the NS bound (4), which is the most interesting
case for us,  $L$ is  restored
  around the NS epoch
 regardless of the value of ${\rm tr}(gg^{\dagger})$ $\cite{foot2}$.

 It may be necessary to further discuss on the parameters
 $\xi$ and $\omega$.  Self interactions among $\chi$'s do not
 change $\rho_{\chi}$, but can increase $n_{\chi}$ through
 the processes like $\chi\chi\ra\chi\chi\chi\chi^{\ast}$.
 The values of $\xi$ and $\omega$  depend on the strength
 of such ``number changing self interaction process" (NCP).
 In our case, NCP's can occur through
  the quartic coupling $\lambda\chi^2\chi^{\ast 2}$
  (loop effects) or through
   nonrenormalizable interactions
 like $\kappa\chi^3\chi^{\ast 3}/\Lambda^2$.
 If the NCP rate is weaker than the expansion rate,
  the average energy of $\chi$
 is roughly equal to that of $\nu$, and then $\xi\simeq\omega\simeq 1$.
In the opposite case,  $\chi$'s would achieve
 a thermal distribution, giving  $\xi\simeq 2$ and $\omega\simeq 1/2$.
 In the subsequent analysis,
  we will simply set $\xi\simeq \omega\simeq 1$ since   as we have
  argued  $Y\leq \xi Y^{\omega}$ and then this choice
   gives  more conservative result  for the $\chi$-production.

Clearly    ${\chi}$'s are mainly produced by the heaviest mass
eigenstate neutrino $\nu$  which has
 the largest Yukawa coupling  $g\equiv{\rm max} \, (g_i)$.
 For interaction terms  in  (2), the processes
which can dominantly produce $\chi$ are as follows:
(A) $\nu\nu\ra \chi$; (B) $\nu\bar{\nu}\ra\chi\chi^{\ast}$;
 (C)
$\nu\nu\ra\chi\chi\chi^{\ast}$ and $\bar{\nu}\bar{\nu}\ra
\chi\chi^{\ast}\chi^{\ast}$.
Before $\chi$'s enter into an equilibrium with
$\nu$'s, particularly when  the NS constraint (4) is satisfied,
 we can safely ignore
the inverse processes annihilating $\chi$'s.
Then the Boltzmann equation describing the evolution
of $\rho_{\chi}$ before the onset of the NS is given by $\cite{ktbook}$
\beq
\frac{d}{dt}\rho_{\chi}+4 H \rho_{\chi}= \sum_I
\int [d X] (2\pi)^4\delta^4(p_i-p_f) E_I |M_I|^2
 f,
\eeq
where $[dX]=\prod_a d^3 k_a/(2\pi^3) 2E_a$  ($a$ runs over particle species
participating in the process),
$E_I$ ($I=A, B, C$) is the total energy of $\chi$'s  in the final state
of the process $I$,
 $|M_I|^2$  is the amplitude squared including
 the symmetry factor for  identical particles,
and finally $f=\prod_i f_i$ for  the phase space distribution functions
$f_i$ of  particles in the initial state.
After a straightforward computation,
this Boltzmann equation  can be cast
into the following form:
\beq
-H\frac{d}{dT}Y=6.7\times 10^{-4}g^2\lambda[
(Z)_A
+(1.6 \frac{g^2}{\lambda}\ln(5/\lambda Z))_B+(10^{-2}\lambda)_C],
\eeq
where   $Z=Y+(g^2/2\lambda)-6(v/T)^2$.
Here the subscripts in the brackets
denote the contributing process. Note that
 the rate of (A) is proportional
to $m^2_{\rm eff}\propto Z$, while that of (B) includes the factor
$\ln ( m_{\rm eff}^2)$.  Of course the above equation
is valid only when $L$ is restored, viz $Z>0$.
At any rate, it indicates that  if $\lambda\geq g^2\ln g^{-2}$,
$\chi$'s are mainly produced
around the NS epoch by the inverse decay process (A) whose rate
 is dominated by the background matter induced piece which  is proportional
 to $Y$.
 However for $\lambda\leq g^2\ln g^{-2}$,
the process  (B) dominates over other processes.
If $\lambda \gg g$, the process (C) can be important at the very early stage
of $Y\ll 1$.

The Boltzmann equation (8)
 can be integrated to obtain
$Y(T_{NS})$. Applying the NS constraint (4), we then find
 \beq
 \lambda g^2\leq 7.2\times 10^{-19}\ln (1+\epsilon \lambda/g^2),
 \eeq
 where  $\epsilon=30/(\lambda^2/g^2
 -160\ln (g^2/5+0.1\lambda))$.
This gives
\beq
g\leq 9\times 10^{-6} R
\eeq
where $R\simeq 1$ for
 $\lambda\leq 100 g^2$, while $R\simeq  (r^{-1}\ln r)^{1/4}$
 for $r=\lambda/100 g^2\gg 1$.
Note that in the case of $\lambda> 100 g^2$,
 the $\chi$-production
is dominated by the process (A) whose rate is enhanced
by the  factor $\lambda/(g^2 \ln g^{-2})$ compared to the rate of
the process (B). This is the reason why we get
a stronger bound on $g$ in this case.

The implication of our NS result  for $\beta\beta_J$ is clear.
Since  $g$ is the majoron Yukawa coupling constant of the
heaviest mass eigenstate neutrino,  $ g_{ee} \leq g$
and thus any upper limit on $g$ applies also to $g_{ee}$.
Then taking into account possible uncertainties
of our analysis, we can conclude that
 $g_{ee}\simeq 10^{-5}$ for which
 observable $\beta\beta_J$
 is barely possible
  can be consistent with the standard NS  model,
   but only under very unlikely conditions
that the quartic coupling constant $\lambda$
is extremely small ($\leq 100 g^2$) and also
the majoron Yukawa couplings with $\nu_{\mu}$ and $\nu_{\tau}$
do {\it not} exceed that with $\nu_{e}$.
This conclusion is valid  as long as $v$ is  small enough,
\eg \,  less than few hundreds keV,
for  $L$  to be  restored around the NS epoch.

Although widely accepted and very natural,  the standard NS model
is {\it not} a unique  model
explaining observed cosmological data.
There may be other successful  models
in which  the constraint  (4) is not valid any more $\cite{ktbook}$,
implying that the NS limit  on $g$ does not have a strictly
firm foundation.
In this regard, it is still worthwhile  to
consider other implications of observable $\beta\beta_J$ with $g_{ee}
\geq 10^{-5}$, \eg \, for supernova dynamics.
It has already been studied how
majoron-like particles  can
 affect the explosion
 and the subsequent cooling of supernovae $\cite{choi,fuller}$,
but  the results of these studies do not show
 any meaningful implication  for $\beta\beta_J$.
Just after the observation
of SN1987A,
 Kolb and Turner $\cite{kolb}$
derived  a limit on the interactions between
supernova neutrinos and  ``cosmic background majorons" (CBM).
   Since the observed neutrino pulse from SN1987A
 is that of $\nu_1$,  the mass eigenstate neutrino  which is mostly $\nu_e$,
 the relevant Yukawa coupling constant here is $g_1$.
 Although in principle
   the parameter $g_{ee}$ describing $\beta\beta_J$ can be
 significantly different from $g_1$
 (note that $g_{ee}=\sum_{i} U^2_{i e} g_i$),
 it is somewhat natural to assume that neutrino mixing is small
enough for  $g_1$ to be close to $g_{ee}$. In this regard, the
  consideration  of CBM's may provide a useful information
 on  $\beta\beta_J$.

  If $\nu_1$'s  from the  supernova
are scattered off by
 CBM's,  there would  be
 a substantial decrease in the average neutrino energy,
 leading to an effective loss of  detectable flux.
  The  observed data
$\cite{1987a}$  indicates that the mean free path $l$ of
$\nu_1$ through the
CBM's should be  comparable to or greater than the distance
to the supernova, viz
\beq
l\geq 1.7\times 10^{23} \, {\rm cm}.
\eeq
 Using this,  it was found  in ref. $\cite{kolb}$ that
$g_1\leq 10^{-3}$.
However as we will see, one can in fact obtain a
significantly stronger  limit
 unless  $\lambda$ is  significantly less than $10^{-3}$.

Clearly for $g_{ee}$ in the range of
 observable $\beta\beta_J$, \ie \, $g_{ee}\geq 10^{-5}$,
  there was a period in the early
universe when $\rho$ and $J$ were at thermal equilibrium
with neutrinos. Later that, the relic majoron number density
can be increased by the decays of $\rho$ and $\nu$ and also
by the $\nu$ annihilations.
With these observations, we will set the majoron temperature
at present as $T_J\simeq 1.9$ K, which is a somewhat
conservative choice.
Then  the inverse mean free path of $\nu_1$  propagating through
CBM's is given by
\beq
l^{-1}\simeq \frac{\sqrt{2}}{4\pi^2}T_J^3\int_0^{\infty}
dw \, w^2 \frac{1}{e^w-1}\int_{-1}^{1} dz \, (1-z)^{1/2}
\sigma(s),
\eeq
where $\sigma (s)$ is the total
cross section for the reaction $\nu_1 J\ra \nu_1 J$
with the total energy-momentum squared  $s= 2ET_J w(1-z)+m_1^2$.
Here  $E$ ($\simeq 10$ MeV) and $m_1$
denote  the energy and the mass of the incident $\nu_1$.
There are three diagrams responsible for the reaction
$\nu_1J\ra \nu_1J$. Two of them are the Compton-type ones,
while the rest one involves the $\rho$-exchange.
The resulting cross section can be written as
\beq
\sigma (s)=\frac{g_1^4}{16\pi s}[ \, F(s)+y^{-1}G(s) \,]
\eeq
where
\bea
F(s)&\simeq& \frac{5}{2}+\frac{2}{x}-
\frac{2(1+2x)^2\ln (1+x)}{x^2(1+x)}+\frac{4+(1-x)\ln y}{1+x}
 \nonumber \\
G(s)&\simeq& \frac{\ln (1+x)}{x^2}-
\frac{1}{x(1+x)},
\nonumber
\eea
for $y=m_1^2/s$ and $x=s/m_{\rho}^2=s/2\lambda v^2$.
Here $F(s)$ represents the contribution
from the Compton-type  diagrams, $y^{-1}G(s)$
is from the $\rho$-exchange diagram.

 If $\lambda$ is so small,  \eg  \, $\lambda\ll
g_1\sqrt{s}/v$,
that the $\rho$-exchange diagram  can be ignored,
we have  $\sigma\simeq g_1^4 \ln (s/m_1^2)/16\pi s$. Then
applying the supernova constraint (11) yields  $g_1\leq 10^{-3}$
as was obtained in ref $\cite{kolb}$.
However in the other case of $\lambda\gg g_1\sqrt{s}/v$,
 the cross section is largely enhanced
by the $\rho$-exchange diagram, allowing
 us to obtain a more stringent limit on $g_1$.
 Our results summarized in fig. 1 shows that,
unless $\lambda\leq 10^{-3}$, some
 portion of the parameter region giving $g_1\leq 3\times  10^{-4}$
 is ruled out by the supernova constraint (11).
As was mentioned, if neutrino mixing is small, so that
 $g_{ee}\simeq g_1$,
 this supernova result can be
 directly applied to $\beta\beta_J$.

To conclude, we find that the standard nucleosynthesis model
strongly constrains the coupling constants in majoron models
in which  observable ``majoron emitting $\nu$-less double beta
decay" ($\beta\beta_J$) is possible
through a spontaneous lepton number violation
below few hundreds keV.
 Our results directly applies to the Yukawa coupling constant
$g_{ee}$ governing $\beta\beta_J$,
yielding  $g_{ee}\leq 9\times 10^{-6}$.
Here the maximal  value $9\times 10^{-6}$
is allowed  only when (i)
the scalar quartic coupling
is extremely weak, roughly  $\lambda\leq 10^{-8}$,
and (ii)  $g_{ee}$ is rather close to
the majoron Yukawa coupling constant of the heaviest
neutrino species.
We also find that the consideration
of the scatterings between  cosmic background majorons
and  supernova neutrinos  provides a
constraint on the majoron Yukawa coupling constant $g_1$
of the mass eigenstate neutrino which is mostly $\nu_e$.
Compared to the nucleosynthesis constraint,
this supernova constraint is much less restrictive,
but is independent of the validity of the standard nucleosynthesis model.
If $g_1\simeq g_{ee}$ and
$\lambda$ is {\it not} less than $10^{-3}$,
which is somewhat plausible, some part of the parameter region
giving observable $\beta\beta_J$ is excluded by the supernova data.

\vskip 1.5cm
\noindent
This work is supported in part by  KOSEF through
CTP at Seoul National University.

\vfill\eject

\newpage
\input prepictex
\input pictex
\input postpictex
\begin{figure}
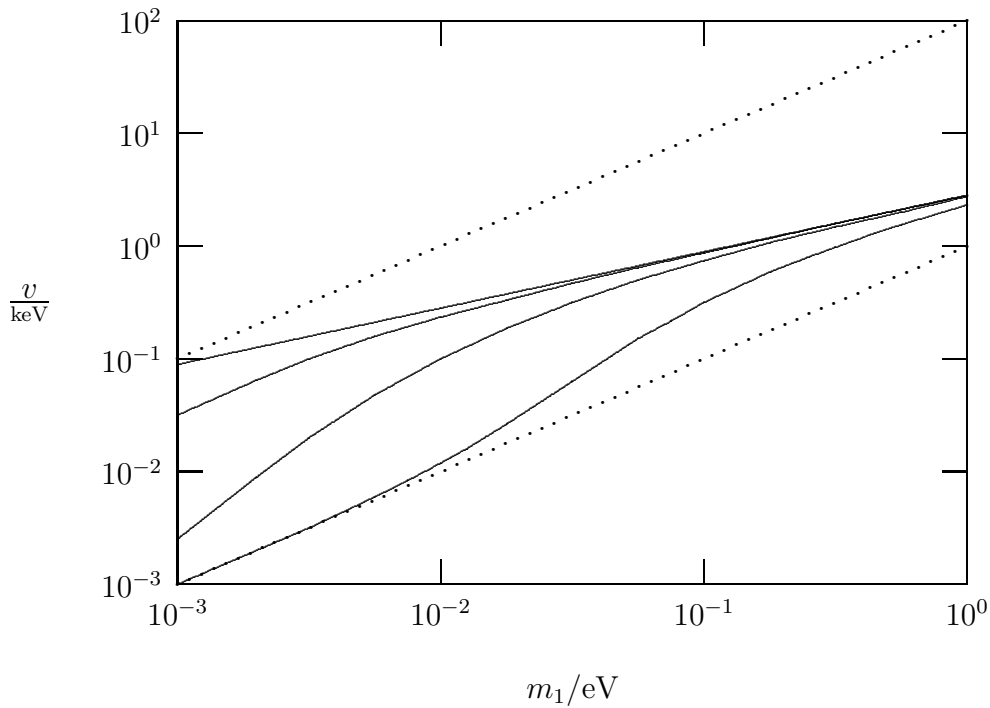
\hskip5mm
\beginpicture
\setcoordinatesystem units <35mm,15mm>
\setplotarea x from -3 to 0, y from -3 to 2
\axis bottom label {$m_1$/eV}
      ticks in withvalues $10^{-3}$ $10^{-2}$ $10^{-1}$ $10^{0}$ / quantity 4 /
\axis left label {$\displaystyle v\over{\rm keV}$}
      ticks in withvalues $10^{-3}$ $10^{-2}$ $10^{-1}$ $10^0$  $10^1$  $10^2$
/ quantity 6 /
\axis right ticks in quantity 6 /
\axis top ticks in quantity 4 /
\plot -3 -1.05 -2 -0.55 -1 -0.05 0 0.45 /
\plot -3    -1.5 -2.7  -1.19 -2.5  -1.0 -2.4  -0.92 -2.25 -0.8
-2    -0.63 -1.75 -0.48 -1.5  -0.33 -1.25 -0.19 -1    -0.06 -0.5   0.2 0
0.45 /
\plot -3     -2.6 -2.7   -2.05 -2.5   -1.7 -2.25  -1.32 -2     -1 -1.75  -0.73
-1.5   -0.5 -1.25  -0.3 -1     -0.13 -0.75   0.031 -0.5    0.17 -0.25   0.3 0
    0.44 /
\plot -3     -3.00 -2.5   -2.5 -2.25  -2.225 -2.125 -2.08 -2     -1.925 -1.9
-1.796
-1.75  -1.58 -1.5   -1.2 -1.35  -0.972 -1.25  -0.82 -1.125 -0.665 -1     -0.5
-0.75  -0.23
-0.6   -0.095 -0.5   -0.01 -0.35   0.115 -0.25   0.19 0       0.365 /
\setplotsymbol ({.}) \plotsymbolspacing=5pt
\setlinear \plot -3 -3 0 0 / \plot -3 -1 0 2 /
\endpicture
\vskip10mm
\caption{ Supernova constraint depicted on the plane of the $L$
breaking scale $v$ and the electron neutrino mass $m_1$.
The solid curves from bottom to top are obtained by applying
the condition (11) for $\lambda=10^{-3}, 10^{-2}, 10^{-1}$, and 1 respectively,
and the regions below them are excluded.
The two dotted lines  correspond to $g_1=10^{-3}$ and $10^{-5}$.  }
\end{figure}


\begin{thebibliography}{99}
\bibitem{foot1}
In ref. $\cite{cline}$, a majoron model in which $L$
 is not spontaneously broken  was considered to  raise up
 $v$  up to few hundreds  MeV.
\bibitem{enqvist}
Note that once $L$ is restored,
the scheme considered by
K. Enqvist {\it et al}., Phys. Rev. Lett. {\bf 68}, 744 (1992)
to  relax the NS constraint (4) through
 $\nu_{\mu},\nu_{\tau}\ra\nu_e J$
can not be applied.
\bibitem{foot2}
In fact $L$ is restored for all $T\geq T_{NS}$
for $Y$ satisfying  the Boltzmann
equation (8) with the b.c. $Y(T_{NS})=0.3$.
\end{thebibliography}
\end{document}